\documentclass[twocolumn,twoside,preprintnumbers,amsmath,amssymb,showkeys]{revtex4}
\usepackage{epsfig}
\usepackage{graphicx}

\usepackage{fancyhdr}

\usepackage{pslatex}

\begin{document}

\newcommand{\slow}{$\sqrt{s_{NN}}$ = 130~GeV }
\newcommand{\shigh}{$\sqrt{s_{NN}}$ = 200~GeV }

\title{Jets and Parton Energy Loss at RHIC: Experimental Status}

\author{John Lajoie$^1$}

\affiliation{$^1$ Iowa State University, Department of Physics and Astronomy, Ames, Iowa 50021}

\begin{abstract}
Experimental evidence from RHIC indicates that matter having an energy density far in excess of the value required for the creation of a deconfined phase 
is produced in ultrarelativistic Au+Au collisions at $\sqrt{s_{_{\rm NN}}}$=200~GeV. This matter thermalizes rapidly, is strongly interacting, and displays 
hydrodynamic properties akin to a fluid with very low viscosity. Studies of the interaction of hard scattered partons with this matter provide an important 
probe of its properties.  In fact, studies of the suppression of high-p$_T$ hadrons have been one of the key arguments used by the RHIC experiments that the 
matter created is something fundamentally new and exciting. I review the current experimental picture that has emerged from the data collected at RHIC, with an 
emphasis on the effort to extract quantitative measurements from the suppression of high-$p_T$ hadrons.  


\end{abstract}
\maketitle

\thispagestyle{fancy}

\setcounter{page}{1}

\section{Jets As Partonic Probes}

The hard scattering of partons in the initial state of an ultrarelativistic heavy-ion collision holds great promise as a probe of
the hot, dense matter created in these collisions. The initial hard scatterings occur very early in the collision process ($\tau <$ 1 fm/c)
and are sensitive to the evolution of the surrounding, color-charged medium primarily through energy loss mechanisms (either collisional
or radiative)\cite{white_papers}. These hard-scattered partons are not observed by themselves in the final state, however, as they fragment into collimated ``jets'' of 
color-singlet hadrons, clustered around the initial parton direction. Experimentally, jets can be identified and studied through the correlation of hadrons 
in the final state, or through the single inclusive spectra of high-$p_T$ particles. In this contribution I will limit the discussion to what we 
have learned about the properties of the medium created in heavy ion collisions through the use of single inclusive spectra. The properties
of the medium that can be inferred through particle correlations are discussed in another contribution to this conference \cite{Olga}. 

\subsection{Hard Probes and pQCD}

In order to use hard scattering as a probe of nuclear matter we need to have a solid understanding of three main components: initial state effects, 
the parton-level cross section, and final state effects (exclusive of the effect of the matter itself). In the collinear factorization ansatz, the cross section
for the production of high-$p_T$ particles can be written as:

\begin{eqnarray}
\lefteqn{\frac{d\sigma^{h}_{n_{1}n_{2}}}{dyd^{2}p_{T}} = } \nonumber \\
& & K\sum_{abcd}\int dx_{a}dx_{b}f_{a/n_{1}}(x_{a},Q^{2})f_{b/n_{2}}(x_{b},Q^{2}) \times \nonumber \\
& & \frac{D_{h/c}(z_{c},Q^{2})}{\pi z} \frac{d\sigma}{d\hat{t}}(ab\rightarrow cd).
\end{eqnarray}

Initial state effects include the distribution of partons in the initail nuclei, as represented by the function $f_{a/n}(x_{a},Q^{2})$. 
For example, a common initial state effect is the multiple scattering of partons prior to the hard scattering event, which results in 
a significant additional transverse momentum of the final state hadrons (the "Cronin effect" \cite{cronin}). 
These effects can be identified through the use of collisions like d+Au collisions at RHIC, 
where the final state effects are expected to be small. In a similar way, our understanding of the partonic level cross section 
$\frac{d\sigma}{d\hat{t}}(ab\rightarrow cd)$ can be 
explored through a comparison between proton-proton collisions and pQCD calculations. In using jets as a probe of the QCD medium, this 
can be thought of as ``calibrating'' the production of these probes in the initial phase of these collisions. Final state effects, primarily energy 
loss as the parton traverses the medium, can be thought of as effectively modifying the fragemtation function $D_{h/c}(z_{c},Q^{2})$, which 
determines the statistical probability of obtaining a hadron $h$ from the fragmentation of a parton $c$, where the hadron has a longitudidinal 
momentum fraction $z_{c}$ of the inital parton. Fragmentation functions have been studied and determined in the relatively clean final states of $e^{+}e^{-}$
collisions. Final state effects such as energy loss can be expected to lower the energy scale of the fragmentation function, as well as increase the 
probability of finding soft hadrons resulting from gluon emission. 

An important first question to ask is whether or not pQCD is applicable at all at RHIC energies. This has been addressed by both the STAR and PHENIX 
collaborations by comparing pQCD calculations with inclusive $\pi^{0}$ and jet spectra in proton-proton collisions \cite{PHENIX_pi0_pQCD, STAR_jets_pQCD}. 
The good agreement between these pQCD
calculations and experimental data indicates the pQCD is broadly applicable at RHIC, and that we have a theoretical handle on the initial state 
(at least in proton-proton collisions). In addition, the demonstration of $x_T$ scaling at RHIC energies in Au+Au collisions confirms that high-$p_T$ 
hadrons originate from hard scatterings in nuclear collisions as well \cite{PHENIX_xT}.   

\subsection{The Nuclear Modification Factor}

In order to understand the effect of hot, dense matter on the production of high-$p_T$ particles, we construct a ratio 
known as the ``nuclear modification factor'' $R_{AA}$:

\begin{equation}
R_{AA}(p_{T}) = \frac{dN^{AA}/dp_{T}}{<N_{binary}>dN^{pp}/dp_{T}}.
\end{equation}

The number of binary collisions for a given centrality class $<N_{binary}>$ is estimated using 
a Glauber model of the nuclear overlap. This overlap is a function of the centrality of the collision, defined as a fraction of the
total inelastic cross section. For example, the ``0-10\%'' most central collisions are the 10\% of collisions, averaged over impact parameter, 
that have the most nuclear overlap. In Au+Au collisions at \shigh $<N_{binary}>\sim 900$. In the niave limit that a nucleus-nucleus collision can be though of
as nothing more than a superposition of independent nucleon-nucleon collisions, the nuclear modification faction at high-$p_{T}$ should be unity. 

\begin{figure}[!htb]
\centerline{\includegraphics*[scale=0.53]{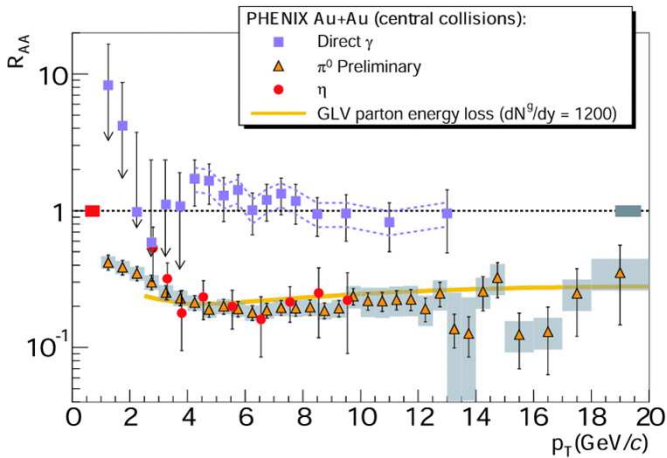}}
\center{\caption{$R_{AA}$ for $\pi^{0}$ and $\eta$ mesons, as well as direct photons, as measured by the PHENIX collaboration.}\label{raa}}
\end{figure}

The measured nuclear modification factor for $\pi^{0}$ and $\eta$ mesons, as well as direct photons from the PHENIX collaboration is shown in Figure~\ref{raa}. 
There is a significant suppression of high transverse
momentum $\pi^{0}$ and $\eta$ mesons observed in Au+Au collisions, consistent with substantial energy loss of the scattered partons. A similar
suppression is observed for charged particles as well as data taken at \slow \cite{phenix_highpt}. In contrast, direct photons show no suppression. This is because the photon 
contains no net color charge, and therefore is able to escape the medium with interacting strongly. The fact that the $R_{AA}$ for direct photons is $\sim$~1 
is another indication that initial state effects are well under control in Au+Au collisions, and that the initial rate of hard scatterings
is properly ``calibrated'' by proton+proton measurements.

\begin{figure}[!htb]
\centerline{\includegraphics*[scale=0.8]{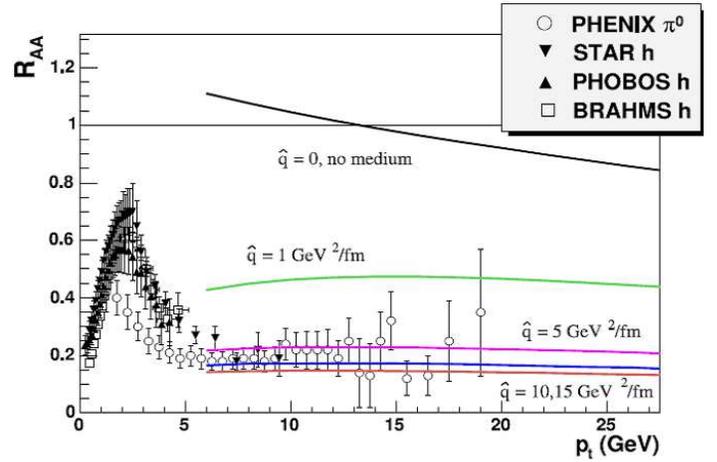}}
\center{\caption{Model fits to the nuclear modification ratio $R_{AA}$ for $\pi^{0}$ mesons and charged hadrons in the BDMPS formalism, 
which fits the data as a function of a 
transport parameter $\hat{q}$. Note that because the suppression is so large, $R_{AA}$ does not offer a great deal of discrimination over a
wide range of $\hat{q}$. Figure from \cite{salgado}.} \label{fig:salgado}}
\end{figure}

Models of energy loss that incorporate the 
expansion of the system indicate the energy loss in the matter created in heavy-ion collisions at RHIC may be as much as fifteen times
larger than the energy loss of a comparable parton in ordinary nuclear matter.  Since the energy loss is 
proportional to the gluon density, this implies the gluon density created in a heavy ion collision is more than an order of magnitude larger than 
in cold nuclei. However, the two leading models for the energy loss of light quarks both different approaches to model the energy loss by gluon radiation. 
The BDMPS model \cite{salgado} characterizes the energy loss by a transport parameter, 
$\hat{q}$, which is the average momentum transfer squared imparted to the parton per unit length of the medium transversed. Formally, this parameter enters
via an integral over the spectrum of emitted gluons, which is related to the medium density of color charges. The value of this parameter, 
averaged over medium expansion, that best fits the data is $\hat{q}$ between 5 and 15 $GeV^{2}/fm $ (see Figure~\ref{fig:salgado}). 
In the GLV model \cite{GLV} the energy loss is
characterized by the initial medium density of gluons $dN_{G}/dy$. The observed suppression in Au+Au collisions is best fit by a value 
$dN_{G}/dy \simeq 1000$ gluons per unit rapidity. It is tempting to attempt to interpret these model parameters as directly related to the properties of the medium. 
This interpretation is called into question, however, by the fact that the authors of these models do not agree on the correspondence 
between $\hat{q}$ and $dN_{G}/dy$. For example, one of of the authors of the GLV model claims the correspondence \cite{GLV_corr}:

\begin{equation}
\frac{dN_{G}}{dy} = 900 \Leftrightarrow \overline{\hat{q}} = 0.35 - 0.85 GeV^{2}/fm
\end{equation}

(where the $\overline{\hat{q}}$ average is done over a 1+1D Bjorken expansion) while the the correspondence claimed by a BDMPS author is \cite{BDMPS_corr}: 

\begin{equation}
\overline{\hat{q}} = 5 GeV^{2}/fm \Leftrightarrow \frac{dN_{G}}{dy} = 900.
\end{equation}

 While both models provide a reasonable description of the data, 
the disagreement between the two approaches makes it questionable to interpret either the BDPMS opacity $\overline{\hat{q}}$ of GLV initial gluon density $dN_{G}/dy$ as measurements
of the physics properties of the QCD medium created at RHIC. 

\begin{figure}[!htb]
\centerline{\includegraphics*[scale=0.6]{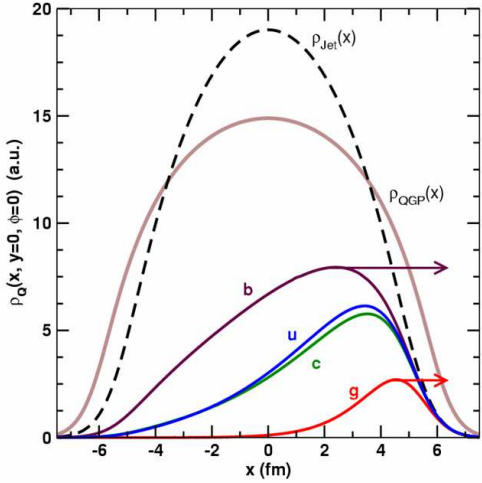}}
\center{\caption{A model calculation showing the the differential character of different flavor probes of the medium. In this simple model, 
partons with $p_{T}=15$ GeV/c orginate in the medium $\rho_{QGP}$ with initial distribution $\rho_{Jet}$ and propagate to the right of the figure. The distribution of surviving jets 
is shown for u,c,b quarks as well as gluons. Note that the b quark, in particular, can provide information about the inner regions of the created medium. 
Figure from \cite{wicks}.}\label{fig:wicks1}}
\end{figure}

In addition to uncertainties in the interpretation of the model calculations, the fact that both models rely on radiative energy loss only has
recently been called into question \cite{collision_pap1}, as will be discussed in the next section. Finally, it is important to note that the large suppression observed
in high-$p_T$ hadrons implies a very large energy loss for light quarks. This in turn implies that the energy loss measurements we make with light quarks
are biased to the surface of the produced medium, and are not necessarily probing the medium in bulk. Ideally, we would like a set of differential probes
that are sensitive to the strongly interacting matter at different effective depths, as shown in Figure~\ref{fig:wicks1}.  
 
The simple model calculations in Figure~\ref{fig:wicks1} make apparant that differential probes of the medium are required, and this 
aspect can be achieved by combining light quark energy loss measurements with measurements of the energy loss of charm and botto quarks. A combination
of such probes will allow better characterization of the QCD matter in bulk, and minimize the surface bias of the light quark data. 

\section{Heavy Quarks}

The measurement of the production of heavy quarks is an exciting probe of heavy-ion collisions and a strong focus of
the future of the RHIC physics program. The energy loss of heavy quarks is suppressed relative to light quarks through interference
of the forward gluon emission amplitudes, resulting in a modified gluon emission spectrum $dI/d\omega$ \cite{annu_rev_bdmps}:

\begin{equation}
\omega \frac{dI}{d\omega}|_{HEAVY} = \frac{\omega \frac{dI}{d\omega}|_{LIGHT}}{1+(\frac{m_Q}{E_Q})\frac{1}{\theta^{2}}} 
\end{equation}

(the so-called ``dead cone'' effect). However, in order to characterize the suppression of heavy quarks in nucleus-nucleus collisions an overall 
knowledge of the total yield in proton-proton 
collisions at RHIC energies is required. At the present time the measurement of heavy quarks at high-$p_T$ is difficult at RHIC because
neither STAR nor PHENIX can identify the displaced vertex associated with the decay of charm mesons. The production of charm quarks is
inferred from the measurement of non-photonic electrons, as described in the next section.
  
\subsection{Inclusive Electrons}

\begin{figure}[!htb]
\centerline{\includegraphics*[scale=0.9]{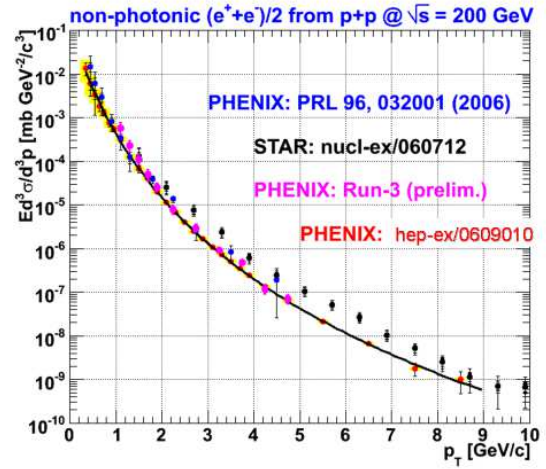}}
\center{\caption{The transverse momentum distribution of single inclusive electrons and positrons in proton-proton 
collisions, as measured by the STAR \cite{STAR_npe} and PHENIX \cite{PHENIX_npe1,PHENIX_new_npe} collaborations. The multiple 
PHENIX datasets are in agreement with each other, within systematic and statistical errors. The line shown is a power-law fit to the most 
recent PHENIX data. Note that the recent STAR data are significantly higher than the PHENIX data at high-$p_T$. }\label{fig:phenix_vs_star}}
\end{figure}

The PHENIX and STAR collaborations have measured the production of single inclusive electrons and positrons from non-photonic sources using 
a meson cocktail subtraction \cite{STAR_npe,PHENIX_new_npe}. The PHENIX experiment cross-checks the cocktail results with a set of special runs 
with an additional photon converter surrounding the beam pipe. Because of the low-mass design of the detector, the PHENIX experiment has a 
substantially higher signal to background than STAR at high-$p_T$. The remaining $e^{+}$ and $e^{-}$ sample is assumed to originate
from the semileptonic decay of heavy quarks. The transverse momentum distribution of electrons and 
positrons from non-photonic sources from PHENIX and STAR is shown in Figure~\ref{fig:phenix_vs_star} for proton-proton collisions at \shigh.
It is important to note that, at the present time, there is a substantial disagreement between the two experiments with the STAR data roughly a factor of
two to three larger than the PHENIX data at high-$p_T$. Until this difference is resolved it will be difficult to properly evaluate the suppression of 
heavy quarks in heavy-ion collisions.  

\begin{figure}[!htb]
\centerline{\includegraphics*[scale=1.0]{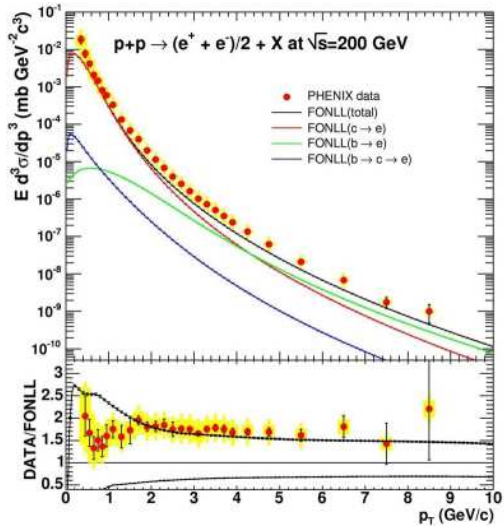}}
\center{\caption{The PHENIX non-photonic electron data compared to FONLL calculations. Figure from \cite{PHENIX_new_npe}. }\label{fig:electrons}}
\end{figure}

\begin{figure}[!htb]
\centerline{\includegraphics*[scale=0.5]{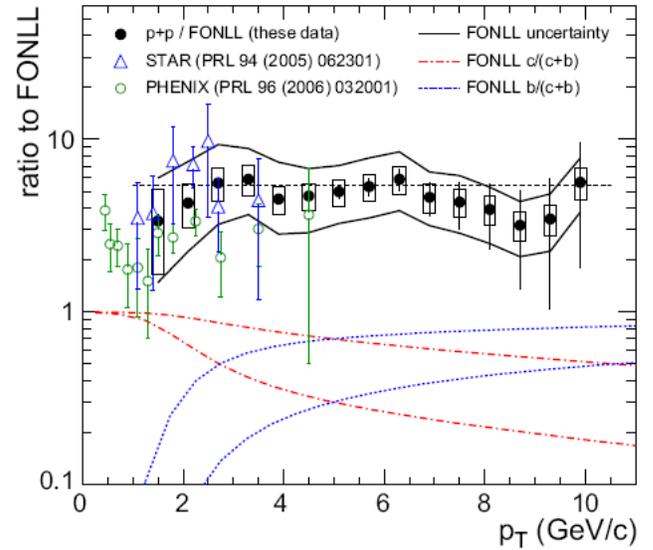}}
\center{\caption{The STAR non-photonic electron data compared to FONLL calculations. Figure from \cite{STAR_npe}.}\label{fig:star_electrons}}
\end{figure}

One interesting comparison that can be made is to compare the non-photonic electron measurements with QCD FONLL (Fixed Order Next To Leading Logartithm) 
calculations, in much the same way we evaluated the applicability of pQCD at RHIC in the beginning of this article. These calculations show a reasonable 
agreement for $J/\Psi$ production in CDF, but the production of $D^0$ mesons is lower than the data by roughly a factor two \cite{CDF_fonll}. 
Figure~\ref{fig:electrons} shows a comparison 
between the PHENIX data and FONLL calculations, while the same is shown for the STAR data in Figure~\ref{fig:star_electrons}.  
It is interesting to note that while the PHENIX data shows a similar behavior to the CDF $D^0$ comparison, namely the data is larger than the FONLL 
calculation by a factor of $\sim$1.5, the STAR data is a factor of $\sim$5 greater than the calculation. In both cases, however, the calculations seem to be in fair
agreement with the shape of the data as a function of $p_T$.

\begin{figure}[!htb]
\centerline{\includegraphics*[scale=0.7]{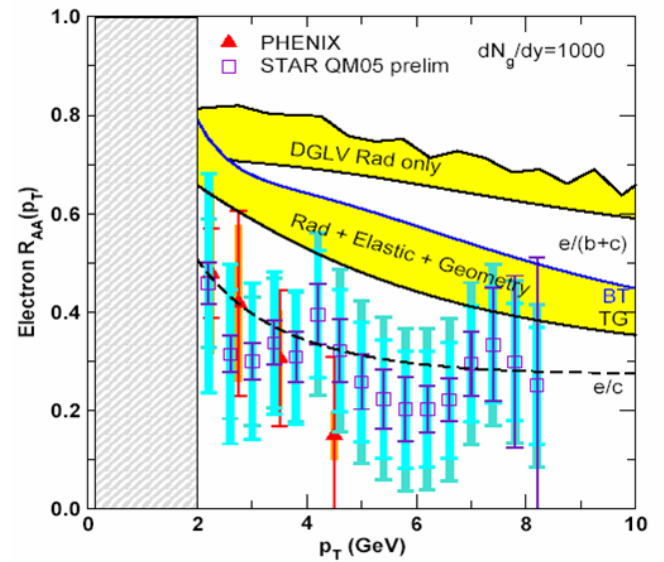}}
\center{\caption{The nuclear modification factor for non-photonic electrons from STAR and PHENIX, shown along with model calculations that
include only energy loss by radiation, or radiation and elastic collisions along with path-length fluctuations through the medium. Figure from \cite{wicks}. }\label{fig:wicks2}}
\end{figure}

The transverse momentum distribution of electrons and 
positrons from non-photonic sources from PHENIX and STAR is shown in Figure~~\ref{fig:wicks2} for 10\% Au+Au collisions at \shigh. Note that both the
PHENIX and STAR data for $R_{AA}$ were calculated using earlier proton-proton reference data. This data is shown compared to model calculations of the energy loss 
of heavy quarks from radiative processes only, and a combination of gluon radiation and elastic collisional losses combined with pathlength fluctuations due to 
collsion geometry. One key point that Figure~\ref{fig:wicks2} makes apparant is that collisional losses were underestimated in models of light quark energy loss, 
and a re-examination of these models is called for. Even so, the range of theoretical values for $R_{AA}$ for heavy quarks is higher than the data from both 
STAR and PHENIX, and only consistent with model calculations that include only charm quarks. This observation, however, may be purely coincidental and if taken
seriously must explain the lower than expected production of bottom quarks at RHIC energies.   

The energy loss of heavy quarks is, at present, an intriguing puzzle in the RHIC data. While further refinement of existing measurements by STAR and PHENIX 
will eventually resolve the discrepancies in the non-photonic electron data, additional detector upgrades will be required before charm and bottom quark 
contributions can be separated and studied independently. Both experiments have upgrades planned (the PHENIX VTX and STAR Heavy Flavor Tracker) that will add
additional high-resolution tracking very close to the beam pipe. These upgrade detectors, combined with expected improvements in luminosity between RHIC I and RHIC II, 
will enable high statistics studies of the propagation of both charm and bottom quarks in the QCD medium of heavy-ion collisions.  

\section{Conclusions}

This initial work at RHIC studying the suppression of high-$p_T$ hadrons in ultrarelativistic heavy-ion collisions is one of the key pieces of
evidence to demonstrate that the matter created in these collisions is something fundamentally new, and very exciting. Model calculations of 
the suppression based on the radiative losses of light quarks as they traverse a colored medium have demonstarted good agreement with the data for different colliding
systems, centralities and energies. However, they have significant shortcomings. The leading model calculations - BDMPS and GLV - differ in the greatly when
trying to draw a correspondence between the parameters of these models, which calls into question the wisdom of interpreting these parameters as representative of 
physical properties of the medium. In addition, it now apprears that collisional losses cannot be fully neglected even for light quarks, and further refinement of the
energy loss models, at least at low-$p_T$, is necessary. Finally, studies of light quark energy loss are inherently biased towards the surface of the created medium
because their interactions with the QCD matter are so strong. 

Studies of heavy quark energy loss offer a less strongly interaction probe of the medium, mainly because of the suppression of forward gluon radiation 
due to the ``dead cone'' effect. Using the charm and bottom quarks as probes will allow an investigation of the medium along a greater pathlength and can potentially
yield additional information about the character of the QCD matter. Early studies at RHIC have made use of the detection of non-photonic electrons to study 
the production of heavy quarks, but current measurements from the STAR and PHENIX collaborations are in disagreement about non-photonic electron measurements 
in proton-proton collisions. It will be difficult to draw conclusions about suppression in heavy ion collisions until this disagreement is resolved. In the future, 
detector upgrade to PHENIX and STAR and luminosity upgrades at RHIC will allow systematic studies of heavy quark energy loss and the explicit separation of 
charm and bottom quark contributions, completing a full set of differential parton probes that can be used to study the hot, dense matter created in 
ultrarelativistic heavy ion collisions.

\hspace{0.25in}  

\noindent{\bf Acknowledgements} 

The author would like to thank the organizers for the invitation to participate in this conference. This work was supported by a grant from the US Department 
of Energy DE-FG02-92ER40692.

\end{document}